\documentclass[aps,preprint,preprintnumbers,showpacs,superscriptaddress,groupedaddress,nofootinbib]{revtex4}  % for double-spaced preprint
\usepackage{graphicx}
\usepackage{epsfig}
\usepackage{amsfonts}

\usepackage{amsmath}
\usepackage{amssymb}
\usepackage{mathrsfs}
\usepackage{graphicx}
\usepackage{float}
\usepackage[breaklinks=true,colorlinks=true]{hyperref}

\hypersetup{colorlinks=true,citecolor=blue,linkcolor=blue,urlcolor=blue}

\usepackage{amssymb}

\newcommand{\insertplot}[5]{\begin{figure}
 \hfill\hbox to 0.05in{\vbox to #5in{\vfill
 \inputplot{#1}{#4}{#5}}\hfill}
 \hfill\vspace{-.1in}
 \caption{#2}\label{#3}
 \end{figure}}
 \newcommand{\inputplot}[3]{
 \special{ps: plotfile #1}
}
\newcounter{fig}

\newcommand{\DS}{\displaystyle}

\newcommand{\ee}{\end{equation}}
\newcommand{\eea}{\end{eqnarray}}
\newcommand{\be}{\begin{equation}}
\newcommand{\bea}{\begin{eqnarray}}

\newcommand{\re}[1]{(\ref{#1})}

\newcommand{\gd}{\mbox{\sf gd}}

\usepackage{xcolor}
\definecolor{mygreen}{HTML}{006E28}

\DeclareMathOperator{\arccot}{arccot}

\begin{document}

\preprint{\small Eur.\ Phys.\ J.\ C \textbf{82}, 757 (2022)   
%\ \ [DOI: 10.1140/epjc/s10052-022-10707-0]
\ [\href{https://doi.org/10.1140/epjc/s10052-022-10707-0}{DOI: 10.1140/epjc/s10052-022-10707-0}]
}
% [arXiv:2106.06399]

\title{Remarks on sine-Gordon kink--fermion system:\\ localized modes and scattering}

\author{Vakhid A. Gani}
\email{vagani@mephi.ru}
\affiliation{Department of Mathematics, National Research Nuclear University MEPhI\\ (Moscow Engineering Physics Institute), Moscow 115409, Russia}
\affiliation{%Kurchatov Complex for Theoretical and Experimental Physics,
%National Research Centre ``Kurchatov Institute'',
%117218 Moscow, Russia
%Ploschad' Akademika Kurchatova 1,
%Moscow 123182, Russia\\
Kurchatov Complex for Theoretical and Experimental Physics\\
of National Research Centre ``Kurchatov Institute'', Moscow 117218, Russia}
%\affiliation{Theoretical Department, National Research Center Kurchatov Institute, Institute for Theoretical and Experimental Physics, 117218 Moscow, Russia}

\author{Anastasia Gorina}
\email{nastya.gorina.2931@gmail.com}
%\affiliation{EPAM Systems, USA}
\affiliation{Osmana\u{g}a Mahallesi Vahapbey Sokak No:8
Kadik\"oy, Istanbul 34714, Turkey}

\author{Ilya Perapechka}
\email{jonahex111@outlook.com}
\affiliation{Wargaming Vilnius, UAB, Konstitucijos pr. 21A, LT-08130 Vilnius}
\affiliation{Department of Theoretical Physics and Astrophysics, Belarusian State University, Minsk 220004, Belarus}

\author{Yakov Shnir}
\email{shnir@theor.jinr.ru}
\affiliation{BLTP, JINR, Dubna 141980, Moscow Region, Russia}
\affiliation{Institute of Physics, University of Oldenburg,
Oldenburg D-26111, Germany}

% 1 % \title{\bf Kink-fermion collision dynamics}

% 1 % \author{
% 1 %       {\large   Vakhid~A.~Gani}$^{\dagger}$,
% 1 % {\large A. Gorina}$^{\star}$,
% 1 %   {\large I.~Perapechka}$^{\star}$ and
% 1 %       {\large   Y. Shnir}$^{\ddagger}$
% 1 % \\
% 1 % \\
% 1 % $^{\dagger}${\small Department of Mathematics, National Research Nuclear University MEPhI}\\
% 1 % {\small (Moscow Engineering Physics Institute), Moscow 115409, Russia}
% 1 % \\
% 1 % $^{\star}${\small Department of Theoretical Physics and Astrophysics, Belarusian State University, Minsk 220004, Belarus}
% 1 % \\
% 1 % $^{\ddagger}$ {\small  BLTP, JINR, Dubna 141980, Moscow Region, Russia}
%\\
%{\small
%Department of Theoretical Physics, Tomsk State Pedagogical University, Russia}
% 1 % }

%\date{January 2022}

% 1 % \maketitle

\begin{abstract}
%{\color{blue}
%We study numerically the kink-fermion interactions in a simple 1+1 dimensional toy model, which describes sine-Gordon kinks coupled to the massless Dirac fermions with backreaction. We show that the spectrum of fermionic modes strongly depends on the choice of the coupling, in particular, there are no localized modes for a minimal Yukawa coupling.     
%
%Our analysis of the collisions between the fermions and the kink demonstrates, that the outcome of the collision dynamically depends on the phase of the incoming fermion packet, it results in alternating regimes of positive and negative acceleration of the kink. 
%}
We study numerically the kink-fermion interactions in a 1+1 dimensional toy model, which describes sine-Gordon kinks coupled to the massless Dirac fermions with backreaction. We show that the spectrum of fermionic modes strongly depends on the choice of the coupling, in particular, there are no localized modes for a minimal Yukawa coupling. We analyze the scattering of the fermionic packet by the kink. We demonstrate that the outcome of the collision dynamically depends on the phase of the incoming fermion packet, it results in alternating regimes of positive and negative acceleration of the kink.
\end{abstract}

%\pacs{05.45.Yv, 11.10.Lm, 11.27.+d, 03.50.-z, %02.60.Cb}

%11.10.Lm --- Nonlinear or nonlocal theories and models
%11.27.+d --- Extended classical solutions; cosmic strings, domain walls, texture,\\
%05.45.Yv --- Solitons,\\
%02.60.Cb --- Numerical simulation; solution of equations,\\
%02.30.Jr --- Partial differential equations,\\
%03.65.Pm --- Relativistic wave equations,\\
%03.50.-z --- Classical field theories,\\
%03.65.Ge --- Solutions of wave equations: bound states,\\

\maketitle

%%%%%%%%%%%%%%%%%%%%%%%%%%%%%%%%%%%%%%%%%%%%%%%%%%%%%%%%%%
\section{Introduction}
%%%%%%%%%%%%%%%%%%%%%%%%%%%%%%%%%%%%%%%%%%%%%%%%%%%%%%%%%%

Many non-linear physical systems support particle-like soliton solutions, which represent spatially localized field configurations with various ramifications for non-linear optics, condensed matter physics, theory of superconductivity, cosmology and quantum field theory \cite{Solitons,Dauxois,Vilenkin,Manton:2004tk,Shnir2018}.

One of the simplest examples of solitons is given by the kinks which appear in models in 1+1 dimensions with a potential possessing two or more degenerate minima, see, e.g., \cite{Manton:2004tk,Shnir2018,Vachaspati,Panos}. The kinks are topologically stable, they interpolate between two different vacua of the model. On the one hand, kinks are widely used to describe objects such as domain walls \cite{Vachaspati} or deformations of planar structures \cite{Yamaletdinov.PRB.2017,Yamaletdinov.Carbon.2019}, and, on the other hand, kink solutions arise, for example, when describing phase transitions in some materials \cite{Khare.PRE.2014}. Depending on the potential of the field-theoretic model, its kink can have different properties. In particular, the asymptotic behavior of the kink solution can be power-law, exponential, or super-exponential, for details see, e.g., \cite{Christov.PRD.2019,Belendryasova.CNSNS.2019,Khare.JPA.2019,Kumar.arXiv.2019.Log,Khare.PS.2019,Khare.arXiv.2019.Log}. Moreover, under certain conditions, compact kinks can exist \cite{Bazeia.PLB.2014}.

A separate exciting area of research is the study of the dynamics of kink-antikink and multikink collisions, see, e.g., \cite{Panos,Gani.EPJC.2021}. Among field models having kink solutions, the so-called integrable models stand out separately. An example of integrable model is the classical sine-Gordon model \cite{Panos-sG}. In the sine-Gordon theory with infinitely degenerate vacuum manifold, the complete integrability of the field equations leads to constraints on the dynamics of the kinks via the infinite number of integrals of motion, see, e.g., \cite{babelon}. In particular, the sine-Gordon model possesses an explicit analytical solution describing fully the dynamics of the kink-antikink pair, the collision of the solitons cannot excite the radiation modes, they do not appear in the final asymptotic state.

Models with polynomial potentials, like for example the simple $\phi^4$ model with the potential $V(\phi)=\frac12(1-\phi^2)^2$ having double degenerate vacuum, are not integrable although have kink solutions. Numerical simulations show that in non-integrable models, like the $\phi^4$ theory, the processes of collisions of a kink and an antikink are chaotic \cite{Anninos:1991un,Belova:1997bq.rus,Belova:1997bq.eng,Campbell:1983xu,Goodman:2005,Makhankov:1978rg,Manton:1996ex,Moshir:1981ja,Dorey:2011yw,Gani:2015cda}. As the impact velocity remains relatively small, the kink-antikink pair annihilate via an intermediate oscillating bion state, however, for some values of the impact velocity the kinks may bounce back via reversible energy exchange between the states of perturbative and non-perturbative sectors of the model \cite{Campbell:1983xu,Belova:1997bq.rus,Belova:1997bq.eng,Anninos:1991un,Dorey:2011yw,Gani:2015cda}. Similarly, a complicated pattern of narrow resonance windows arises in the deformed sine-Gordon models \cite{Peyrard:1983rzn,Dorey:2021mdh,Campbell:1986nu,Gani:2017yla}.

The complete integrability of the sine-Gordon theory also can be destroyed by the presence of impurities \cite{Panos-sG,malomed,malomed2,Kivshar.PLA.1985}, modification of the boundary conditions \cite{Arthur:2015mva}, or via coupling with other fields. In particular, fermions can be coupled to the sine-Gordon model via the Yukawa interaction term \cite{Brihaye:2008am}. Remarkably, there are fermion modes trapped by the kink, for certain values of the parameters of the model there exists a localized zero-energy mode. Indeed, the underlying Atiyah-Patodi-Singer index theorem \cite{APS1,APS2} requires one normalizable fermion zero mode per unit topological charge of the background soliton in the scalar sector.

Fermions bound by kinks were considered in many papers \cite{Brihaye:2008am,Jackiw:1975fn,Dashen:1974cj,Chu:2007xh,Liu:2008pi,GaKsKu01.rus,GaKsKu01.eng,GaKsKu02.rus,GaKsKu02.eng,Campos.JHEP.2021}. The typical assumption in most of such considerations is that the the spinor field does not backreact on the soliton. Moreover, only the fermion zero modes were considered in most cases. A different approach to the problem was proposed in our previous works \cite{Perapechka:2018yux,Perapechka:2019dvc,Klimashonok:2019iya,Perapechka:2019vqv} where we reconsidered this problem taking into account both the backreaction of the localized fermions, and the spectral flow of the eigenvalues of the Dirac operator. We observed that the interaction with fermions induces deformations of the kinks, which may resemble the excitations of the internal modes of the scalar configurations \cite{Klimashonok:2019iya}.
% \cite{Klimashonok:2019iya,Manton:1996ex,Manton:2019xiq}

Presence of fermion bound states affects the process of collision of kinks, there are transitions between localized fermion modes during the collision \cite{Gibbons:2006ge,Saffin:2007ja,Campos:2020iov} and radiation of fermions. This raises an interesting question: what happens to the kink during collisions with incoming fermions?

It is known that the influence of an incident scalar radiation on the kink may produce both positive and negative radiation pressure, depending on the form of the scalar potential of the model \cite{Romanczukiewicz:2003tn,Forgacs:2008az}. The effect of negative radiation pressure in the $\phi^4$ theory is related with non-linearity of the model, the excitation of the kink by incoming radiation with some frequency $\omega$, induces the outgoing wave whose frequency is twice of $\omega$, this yields a surplus of momentum created behind the kink \cite{Romanczukiewicz:2003tn,Forgacs:2008az}. On the other hand, the sine-Gordon kink is transparent to the incoming scalar radiation \cite{Forgacs:2008az}.

A main objective of this Letter is to study the effects of dynamical interaction between the sine-Gordon kinks and incoming fermions numerically, taking into account backreaction of the scalar field. Our computations reveal that deformation effects and excitation of the scalar radiation in the process of collision, result in a complicated pattern of chaotic scattering with alternating regimes of positive and negative acceleration of the kink.

This paper is organized as follows. 
Section \ref{sec:Model} recalls briefly the model, here we also discuss the choice of the Yukawa coupling. In Section \ref{sec:Fermions}, we review the types of the fermionic modes localized on the kink, we also present the results obtained by solving the coupled system of equations numerically. Then, in Section \ref{sec:Collisions} we investigate the collision dynamics of a massless fermion and a sine-Gordon kink. We close with our conclusions in Section \ref{sec:Conclusion}.

%%%%%%%%%%%%%%%%%%%%%%%%%%%%%%%%%%%%%%%%%%%%%%%%%%%%%%%%%%%%%%%%%
\section{The model}
\label{sec:Model}
%%%%%%%%%%%%%%%%%%%%%%%%%%%%%%%%%%%%%%%%%%%%%%%%%%%%%%%%%%%%%%%%%%
The 1+1 dimensional field theory we are interested in is defined by the following Lagrangian
\be \label{lag}
\mathcal{L} =
\frac{1}{2}\partial_\mu\phi\partial^\mu\phi +
\bar{\psi}\left[i\gamma^\mu\partial_\mu-m-g\phi\right]\psi - V(\phi)
\, ,
\ee
where $V(\phi)$ is a potential of the self-interacting real scalar field $\phi$, $\psi$ is a two-component Dirac spinor, $m$ is the bare mass of the fermions and $g$ is the Yukawa coupling constant. The matrices $\gamma_\mu^{}$ are $\gamma_0^{}=\sigma_1^{}, \gamma_1^{}=i\sigma_3^{}$ where $\sigma_i^{}$ are the Pauli matrices and $\bar{\psi}=\psi^\dagger\gamma^0$. The fermion field is minimally coupled to the scalar sector via the Yukawa interaction term
\be \label{Yukawa}
\mathcal{L}_{\rm int}^{} = -g \phi \bar \psi \psi\, .
\ee
Note that this form of the coupling of the scalar and spinor fields,  for any values of the parameters of the model, is different from the case of the supersymmetric sine-Gordon system  \cite{Boya:1990fp} with simple prepotential $W=-4\cos(\phi/2)$. Hence, there is no simple relation between bosonic and fermionic spectra of excitations, they are no longer related to each other.

The Euler-Lagrange equations for the model \re{lag} are
\be
\begin{split}
    \label{FieldEquations ScalarFermionField}
    \partial_\mu\partial^\mu\phi+g\bar{\psi}\psi - V^\prime(\phi) = 0\, ,\\
    i\gamma^\mu\partial_\mu\psi-m\psi-g\phi\psi = 0\, .
\end{split}
\ee
The sine-Gordon model corresponds to the periodic potential $V(\phi)=1-\cos \phi$ with infinite number of degenerate vacua $\phi_0^{}=2\pi n$, $n$ is an integer.
%$n\in \mathbb{Z}$.

In the decoupled limit ($g=0$) the static sine-Gordon kink solution interpolating between the vacua $0$ and $2\pi$ is well known,
\be \label{kink}
\phi_{\rm K}^{}(x) = 4 \arctan (e^{x}) = 2 \arcsin(\tanh x) +
\pi = 2\:\gd(x) + \pi \, ,
\ee
where $\gd(x)=2\arctan(e^x)-\displaystyle\frac{\pi}{2}$ is the Gudermannian function \cite{Gradshtein}. Fermion modes can be localized on the sine-Gordon kink after appropriate adjustment of the coupling in the Yukawa interaction term \re{Yukawa}, $g \phi_{\rm K}^{} \to g(\phi_{\rm K}^{}-C)$ \cite{Brihaye:2008am}. Equivalently, the mass of the fermion field can be shifted as $m \to m-g C$, in particular, setting $C=\pi$ restores the reflection symmetry of the kink solution $x\to -x,\;\; \phi\to-\phi$.

Setting $C=\pi$ and making use of the usual parametrization of a two-component spinor
\begin{equation}
\label{ansFer} \psi(x,t) = e^{-i\epsilon t}\left(
\begin{array}{c}
u(x)\\
v(x)
\end{array}
\right),
\end{equation}
where $u(x)$ and $v(x)$ are two real functions, we obtain the following coupled system of static equations:
\be \label{eq2}
\begin{split}
\phi_{xx} + 2g uv -\sin \phi_{\rm K}^{} &=0\, ,\\
u_x+(m-g\pi+g\phi_{\rm K}^{})u&=\epsilon v\, ,\\
-v_x+(m-g\pi+g\phi_{\rm K}^{})v&=\epsilon u\, .
\end{split}
\ee
This system is supplemented with the normalization condition
\begin{equation}\label{eq:norm_cond}
    \int\limits_{-\infty}^\infty \left(u^2+v^2\right)dx = 1\, ,
\end{equation}
thus the configuration as a whole can be characterized by two quantities, the fermionic density distribution $\rho_{\rm f}^{}=u^2+v^2$ and the topological density of the kink $\rho_{\rm t}^{}=\DS\frac{1}{2\pi}\frac{\partial\phi}{\partial x}$.

%%%%%%%%%%%%%%%%%%%%%%%%%%%%%%%%%%%%%%%%%%%%%%%%%%%%%%%%%%%%%%%%%
\section{Localized fermions}
\label{sec:Fermions}
%%%%%%%%%%%%%%%%%%%%%%%%%%%%%%%%%%%%%%%%%%%%%%%%%%%%%%%%%%%%%%%%%%

Assuming that the scalar field background \re{kink} is fixed and setting bare mass of the fermion $m=0$, we can obtain a solution for the localized zero mode of the Dirac equation (up to a normalization factor) \cite{Brihaye:2008am,Bazeia:2017zye}:
\be\label{fermi_zero}
    \psi_0^{}(x) \propto \left(
    \begin{array}{c}
        \exp\left\{-g\left[2x (2\arccot(e^x) +\gd(x)) + 4\:{\rm Ti}_2^{}(e^{-x})\right] \right\}\\
        0
    \end{array}
    \right),
\ee
where ${\rm Ti}_2^{}(y)$ is the inverse tangent integral of $y$. By expanding the functions in powers of $x$ around the center, we can see that the fermion zero mode \re{fermi_zero} can be nicely approximated by a gaussian profile, see Fig.~\ref{fig1} below.

Other localized modes can be found in the same way. The system of two first order differential equations in \re{eq2} can be transformed into two decoupled second order equations for the components $u$ and $v$:
\be\label{eq3}
\begin{split}
-u_{xx}+\left( \left(m-g\pi + 4 g  \arctan (e^{x})\right)^2 - \frac{2g }{\cosh x} \right) u&=\epsilon^2 u\, ,\\
-v_{xx}+\left( (m-g\pi + 4 g  \arctan (e^{x}))^2 + \frac{2g}{\cosh
x} \right) v&=\epsilon^2 v\, .
\end{split}
\ee
They are Schr\"odinger-type equations, for the fermions in the external static background field of the kink \re{kink} with the potential well
\be \label{pot-fermi}
U_\mp(x) = (m-g\pi + 4 g  \arctan (e^{x}))^2 \mp
\frac{2 g }{\cosh x}\, .
\ee
Since $U_\pm(x) \to (m-g\pi)^2$ at $x\to-\infty$, and $U_\pm(x) \to (m+g\pi)^2$ at $x\to+\infty$, in the massless limit $m=0$ the functions $u,v$ decay as $\sim e^{-\lambda |x|}$, where $\lambda=\sqrt{g^2\pi^2-\epsilon^2}$. Further, setting bare mass of the fermion field to zero reduces the system \eqref{eq3} to the form
\begin{equation}\label{eq:equations_with_Hamiltonians}
    \hat{H}_1^{} u(x) = \epsilon^2 u(x)\, , \qquad \hat{H}_2^{} v(x) = \epsilon^2 v(x)
\end{equation}
with the Hamiltonians
\begin{equation}\label{eq:Hamiltonians}
    \hat{H}_1^{} = - \frac{d^2}{dx^2} + U_{-}^{}(x)\, , \qquad \hat{H}_2^{} = - \frac{d^2}{dx^2} + U_{+}^{}(x)\, ,
\end{equation}
where
\be\label{eff-pot}
    U_\mp(x) = 4g^2\: \gd^2(x) \mp 2g\:\frac{d}{d x} \gd(x)\, .
\ee
The operators \eqref{eq:Hamiltonians} can be factorized as
\begin{equation}\label{eq:Hamiltonians_factorized}
    \hat{H}_1^{} = \hat{a}^\dagger \hat{a}\, , \qquad \hat{H}_2^{} = \hat{a}\: \hat{a}^\dagger\, ,
\end{equation}
where the following operators are introduced:
\be \label{ladder_fermi-op}
\hat{a} = \frac{d}{dx} + w(x)\, , \qquad \hat{a}^\dagger = - \frac{d}{dx} + w(x)
\ee
with the superpotential
\begin{equation}\label{eq:superpotential}
    w(x) = 2g~\gd(x)\, .
\end{equation}
It is easy to see that the potentials \eqref{eff-pot} can be expressed via the superpotential \eqref{eq:superpotential} as
\begin{equation}
    U_\mp(x)=w^2(x)\mp w^\prime(x)\, .
\end{equation}
The algebra of the operators \eqref{ladder_fermi-op} is
\be
[\hat{a},~\hat{a}^\dagger] = 2\:w^\prime(x) = \frac{4g}{\cosh x} \, .
\label{alg_a}
\ee

The discrete spectrum of each Hamiltonian $\hat{H}_1^{}$ and $\hat{H}_2^{}$ contains a finite number of levels, depending on the coupling constant $g$. Action of the operators \eqref{ladder_fermi-op} on the eigenfunctions $u_n^{}(x)$ and $v_n^{}(x)$ of the Hamiltonians $\hat{H}_1^{}$ and $\hat{H}_2^{}$ is the following (up to normalization constants):
\begin{equation}
    \hat{a}\: u_{n+1}^{} = v_n^{}\, , \qquad \hat{a}^\dagger v_n^{} = u_{n+1}^{}\, ,
\end{equation}
see, e.g., \cite{Cooper.PhysRep.1995}. The ground state $u_0^{}$ is annihilated by the operator $\hat{a}$,
\be \hat{a}\: u_0^{}(x) = \left(\frac{d}{dx} +2g ~\gd(x)\right) u_0^{}(x) = 0\, ,
\ee
the solution of this equation is the fermion zero mode \re{fermi_zero}. Note that similar operators can be also defined in the case $m\neq 0$.

Let us compare Eqs.~\re{eq3} with the corresponding Schr{\"{o}}dinger-type equation on the scalar excitations of the sine-Gordon kink, $\phi(x,t) = \phi_{\rm K}^{}(x) + \eta(x)\:e^{i\omega t}$:
\be \label{kink-exiced}
    -\eta_{xx} +\left(1-\frac{2}{\cosh^2 x} \right)\eta = \omega^2\eta
\ee
with the P\"oschl--Teller-type potential $U(x)=1-\DS\frac{2}{\cosh^2 x}$. By analogy with \eqref{eq:equations_with_Hamiltonians} and \eqref{eq:Hamiltonians_factorized}, this equation also can be expressed in terms of the operators $\hat{b}$ and $\hat{b}^\dagger$ as
\be \label{ladder_scalar}
    \hat{b}^\dagger \hat{b}\: \eta(x)=\omega^2 \eta(x) \, ,
\ee
where
\be \label{ladder_boson-op}
\hat{b} = \frac{d}{dx} +\tanh x \, , ~\qquad \hat{b}^\dagger = - \frac{d}{dx} + \tanh x\, ,
\ee
and
\be
    [\hat{b},~\hat{b}^\dagger] = \frac{2}{\cosh^2 x}\, .
\ee
The zero mode of the sine-Gordon kink is defined as a state, which is annihilated by the action of the operator $\hat{b}$:
\be
    \hat{b}\:\eta_0^{}(x) = \left( \frac{d}{dx} +\tanh x \right) \eta_0^{}(x) = 0 \, .
\ee
Up to a normalization factor, it looks like
\be\label{scalar_zero}
    \eta_0^{}(x) = \frac{1}{\cosh x}\, .
\ee
The sine-Gordon continuum scalar modes are
\be\label{scalar_cont}
\eta_k^{}(x) = e^{ikx}\frac{ik - \tanh x}{\sqrt{1+k^2}}\, , \qquad k^2 = \omega^2 -1 \, .
\ee

It is noteworthy that the partner Hamiltonian $\hat{b}\:\hat{b}^\dagger$ has the constant potential $U(x)\equiv 1$, which leads to the absence of discrete levels. At the same time, the discrete spectrum of Eq.~\eqref{kink-exiced} contains the only zero mode $\omega_0^{}=0$ with the corresponding eigenfunction \eqref{scalar_zero}.

Clearly, the  fermionic and bosonic operators \re{ladder_fermi-op}, \re{ladder_boson-op} are not directly related, and the effective potential \re{pot-fermi} or \eqref{eff-pot} is different from the reflectionless P\"oschl--Teller-type potential in Eq.~\eqref{kink-exiced}.

Unfortunately, Eqs.~\re{eq3} do not support solutions of the form of a hypergeometric-type function, moreover, even in the limit of zero bare mass $m=0$, the action of the operators \re{ladder_fermi-op} does not provide  closed form solutions of Eqs.~\re{eq:equations_with_Hamiltonians}. Hence, in our analysis we have to implement numerical methods.

\subsection*{Numerical results}

Since an analytical solution of the full system of integral-differential equations \re{eq2}, \eqref{eq:norm_cond} cannot be attained, we solved this problem numerically implementing the method of lines. The spatial derivatives are discretized on an equidistant grid using fourth-order finite difference method and the resulting system of ordinary differential equations is integrated using eighth order Dormand--Prince method.

First, we consider fermionic modes, localized on the fixed background of the sine-Gordon kink, so we adjust the Yukawa coupling as above, $g \phi_{\rm K}^{} \to g(\phi_{\rm K}^{}-\pi)$. Then, in the limit of zero bare mass of the fermions, $m=0$, the effective potential \re{eff-pot} becomes symmetric (the  Gudermannian function $\gd(x)$ is parity-odd). Hence, by analogy with the fermions localized on the $\phi^4$ kink \cite{Jackiw:1975fn,Dashen:1974cj,Chu:2007xh,Klimashonok:2019iya}, there are two types of localized modes with opposite parity with the following boundary conditions at the center:
\be
{\rm {Type ~A}:}~~u_x\biggl.\biggr|_{x=0}=0\, , \quad v\biggl.\biggr|_{x=0}=0\, ; \qquad
{\rm {Type ~B}:}~~u\biggl.\biggr|_{x=0}=0\, , \quad v_x\biggl.\biggr|_{x=0}=0\, .
\ee
In other words, the modes $A_n$ of the type $A$ possess symmetric $u$-component and antisymmetric $v$-component, the modes $B_n$ of the type $B$ possess antisymmetric $u$-component and symmetric $v$-component. Spinor components of each mode possess at least $n$ nodes, the zero mode is labeled as $A_0^{}$.

\begin{figure}[t!]
    \begin{center}
    \includegraphics[trim=50 30 70 30,clip,width=0.4\textwidth]
        {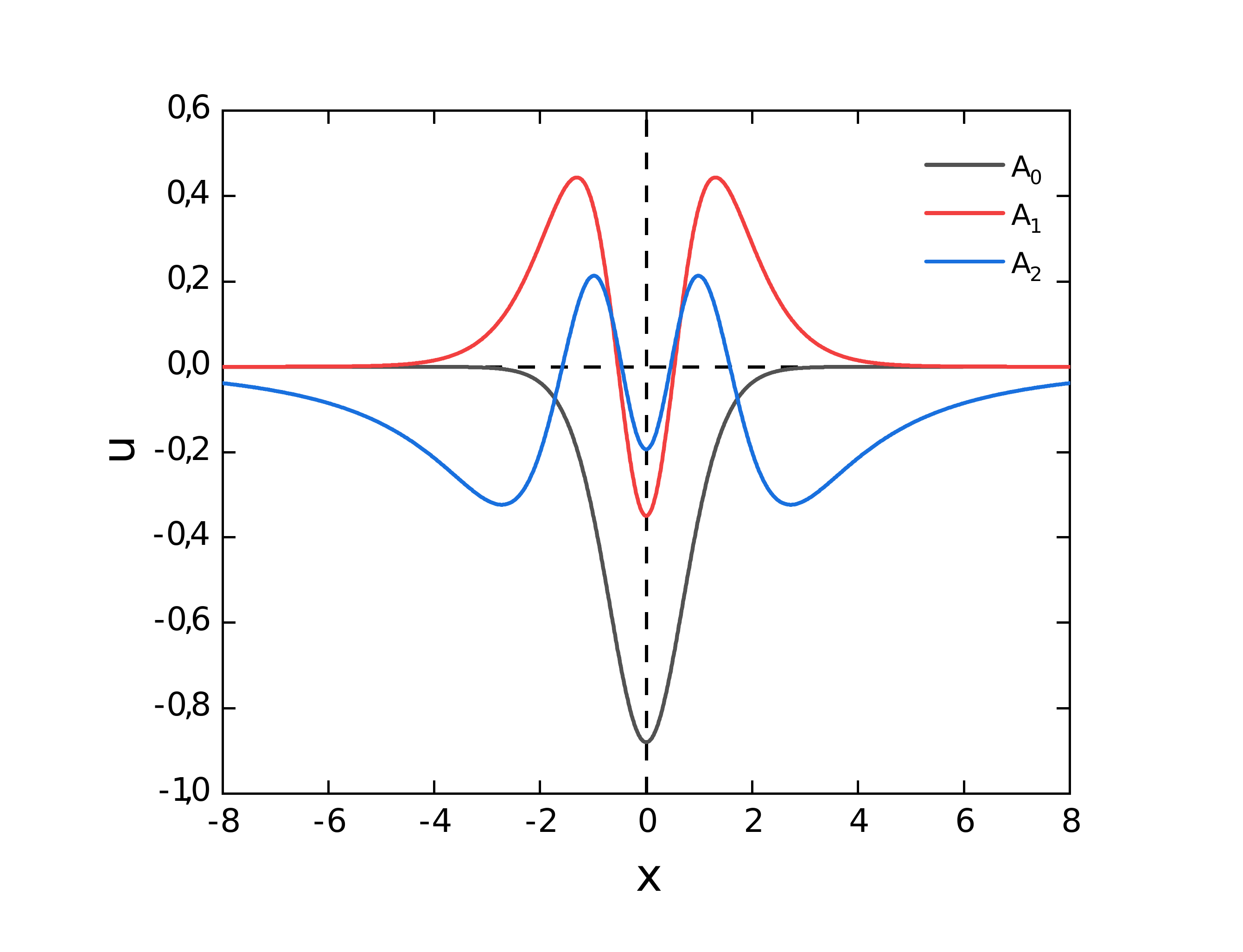}
    \includegraphics[trim=50 30 70 30,clip,width=0.4\textwidth]
        {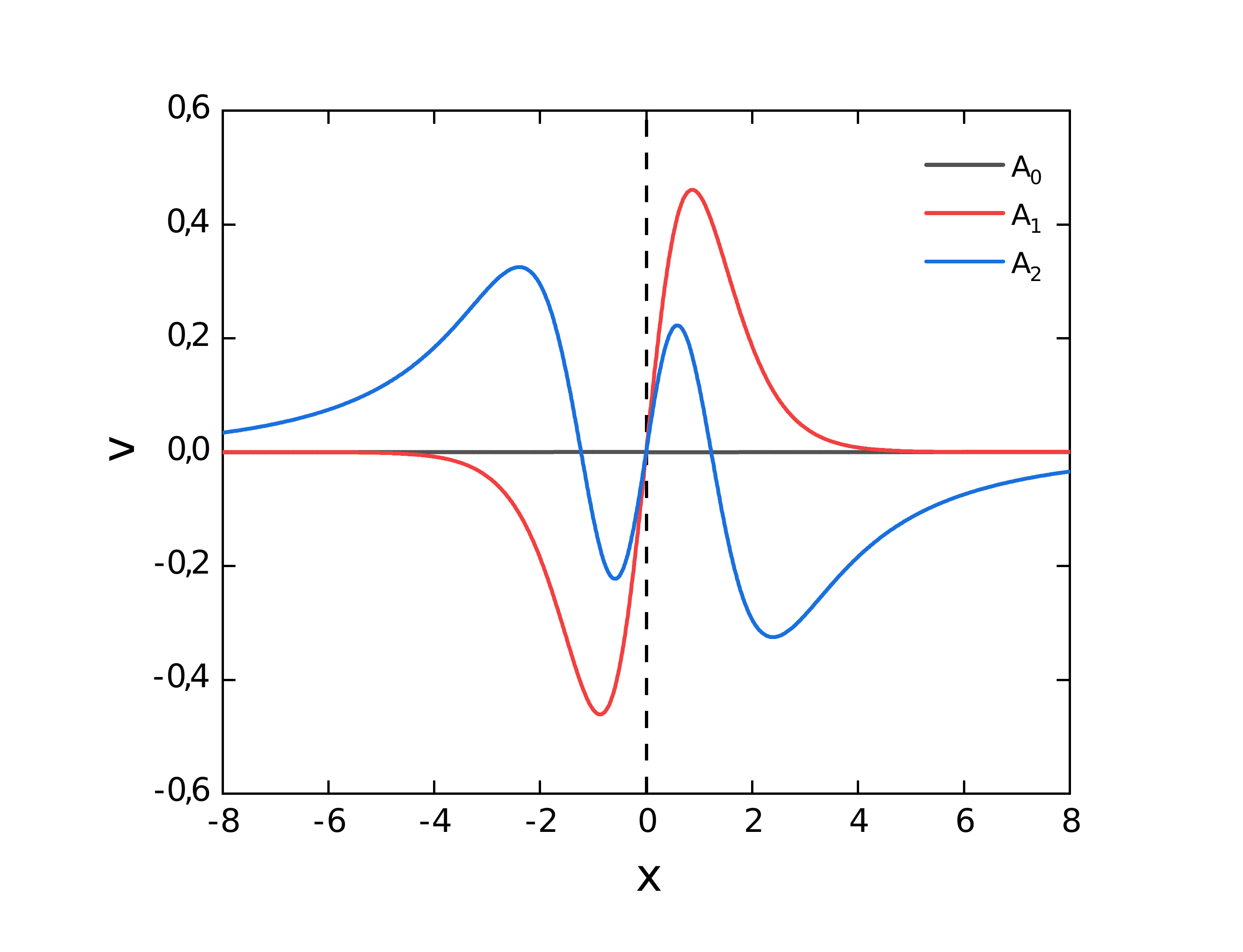}
    \includegraphics[trim=50 30 70 30,clip,width=0.4\textwidth]
        {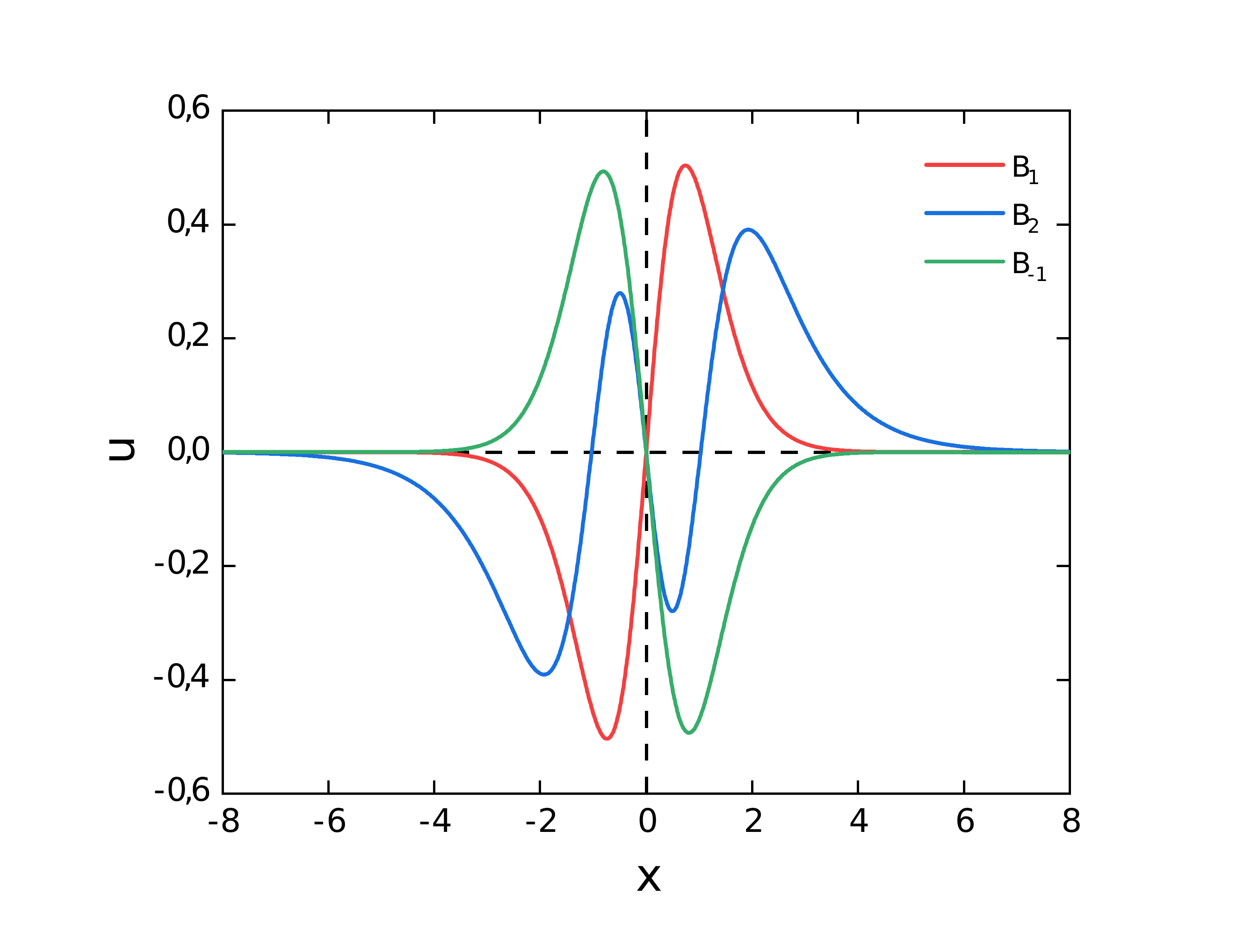}
    \includegraphics[trim=50 30 70 30,clip,width=0.4\textwidth]
        {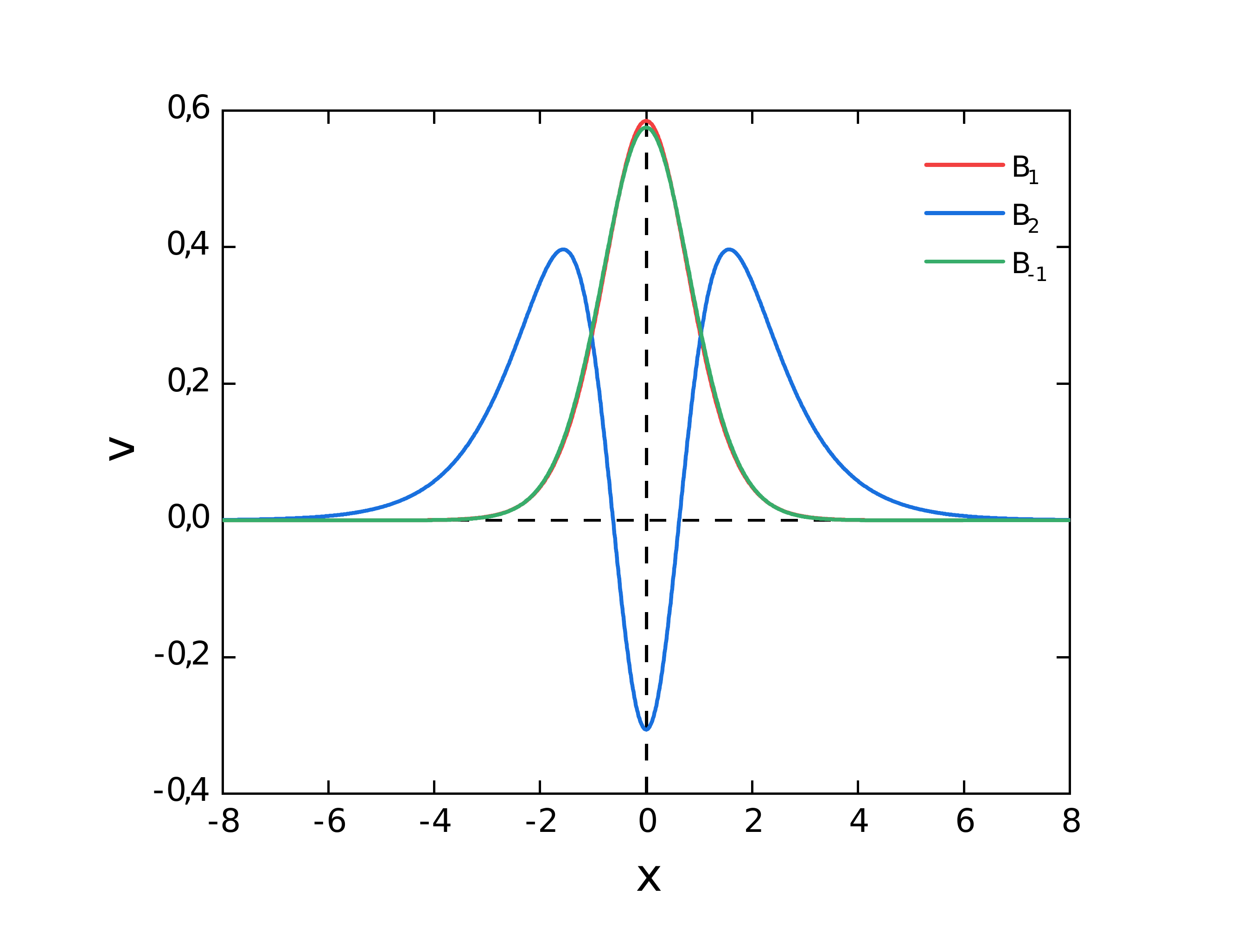}
    \end{center}
    \vspace*{0mm}
    \caption{\small{Components of the localized fermionic modes of the types $A_n$ (upper row) and $B_n$ (bottom row) are plotted as functions of the coordinate $x$ for $m=0$ and $g=1$.}}
    \label{fig1}
\end{figure}

Another similarity with the fermionic modes localized on the $\phi^4$ kink is that for all spinor states, the number of nodes of the component $u$ is one node more than the number of nodes of the component $v$. In Fig.~\ref{fig1},
we display the components of a few modes of both types, localized on the sine-Gordon kink at $g=1$. Here we fix the mass parameter $m=0$ as above.

\begin{figure}[t!]
    \begin{center}
    \includegraphics[trim=50 30 70 30,clip,width=0.6\textwidth]
        {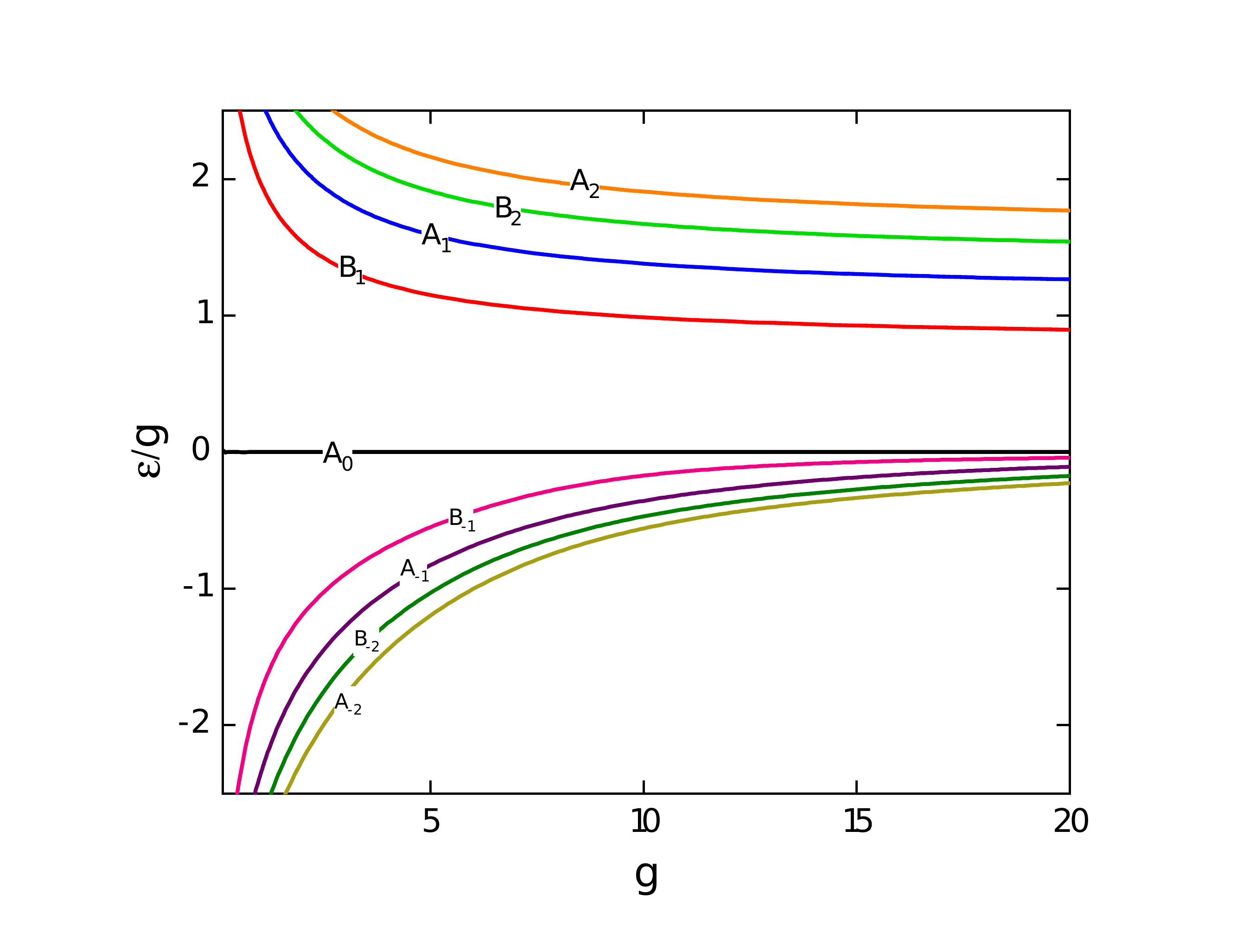}
    \end{center}
    \vspace*{0mm}
    \caption{\small{Normalized energy of the localized fermionic states as a function of the Yukawa coupling $g$
    for first nine modes at $m=0$ with backreaction.}}
    \label{fig2}
\end{figure}

The number of the localized massless fermionic states and their energies depend on the strength of the Yukawa coupling. In Fig.~\ref{fig2}, we plot spectral flow of the first five localized modes of both types.

\begin{figure}[t!]
    \begin{center}
    \includegraphics[trim=50 30 70 30,clip,width=0.4\textwidth]
        {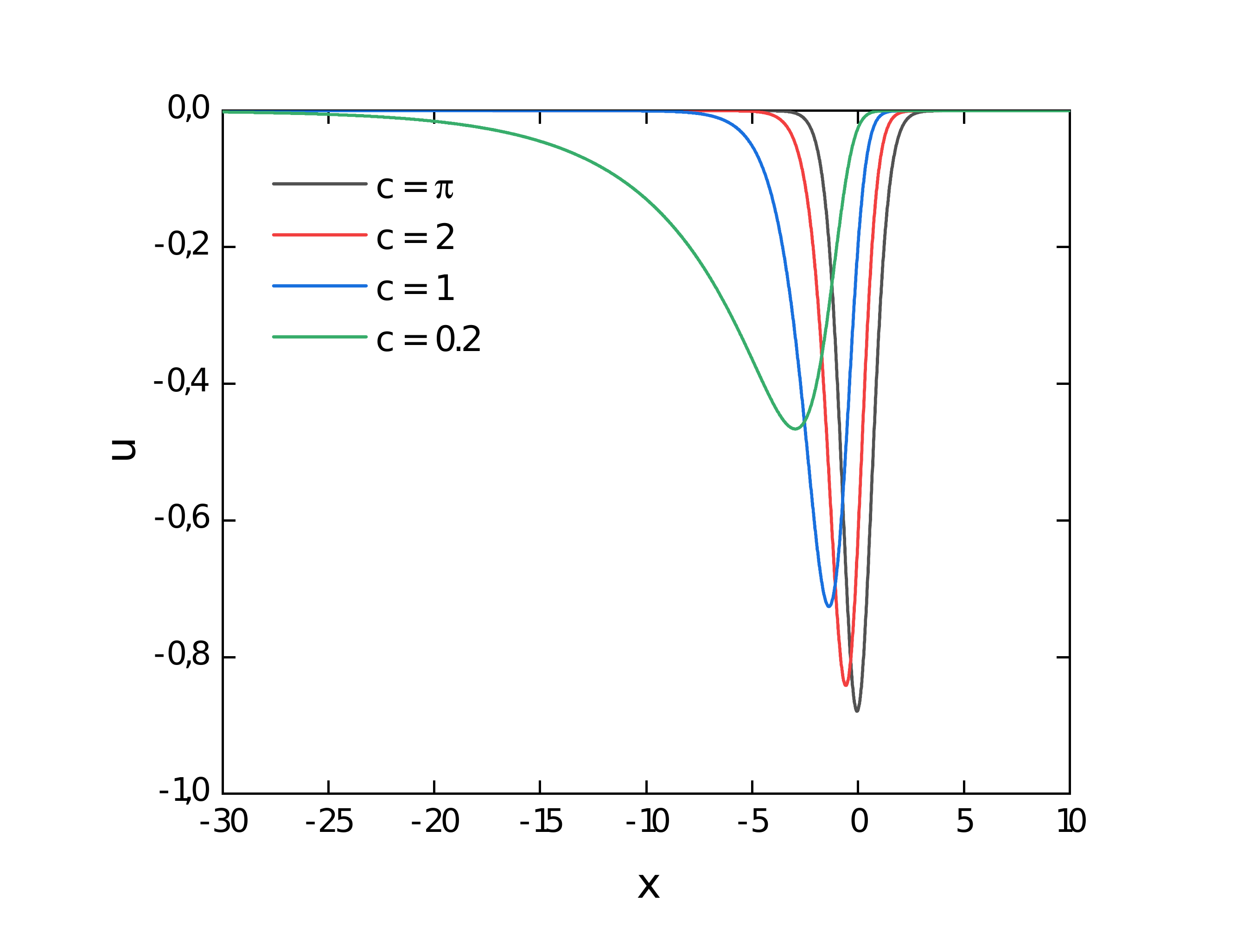}\\
    \includegraphics[trim=50 30 70 30,clip,width=0.4\textwidth]
        {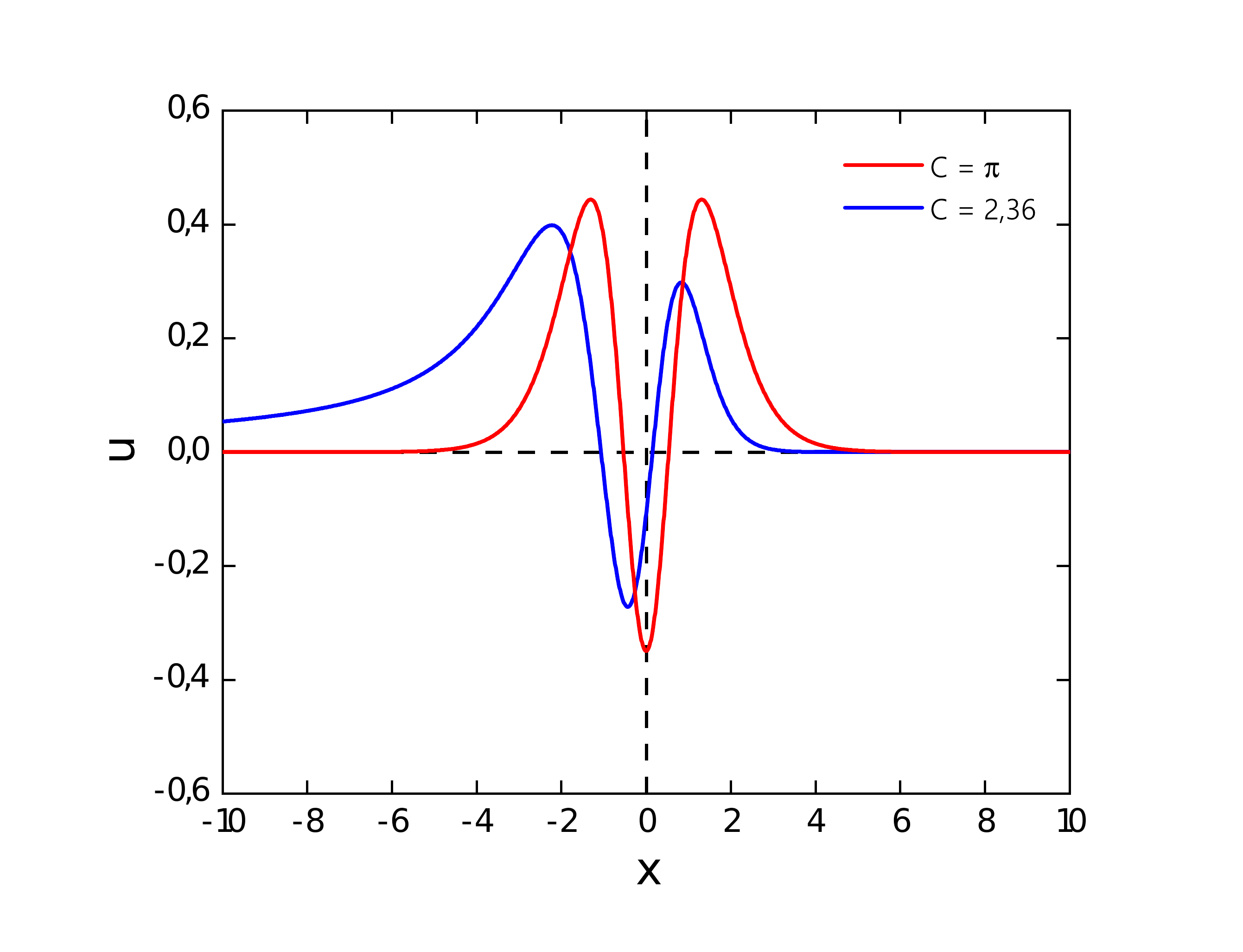}
    \includegraphics[trim=50 30 70 30,clip,width=0.4\textwidth]
        {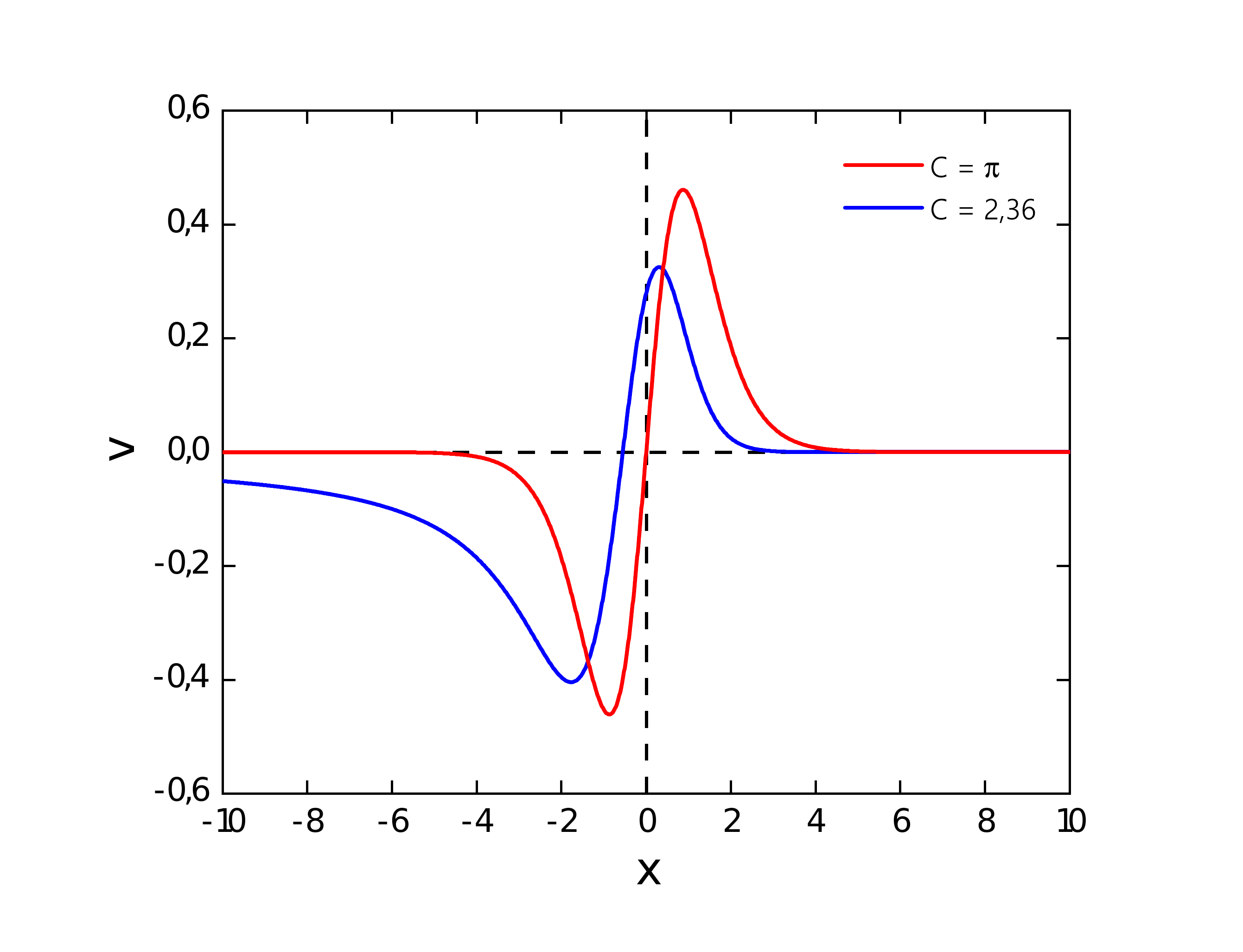}
    \includegraphics[trim=50 30 70 30,clip,width=0.4\textwidth]
        {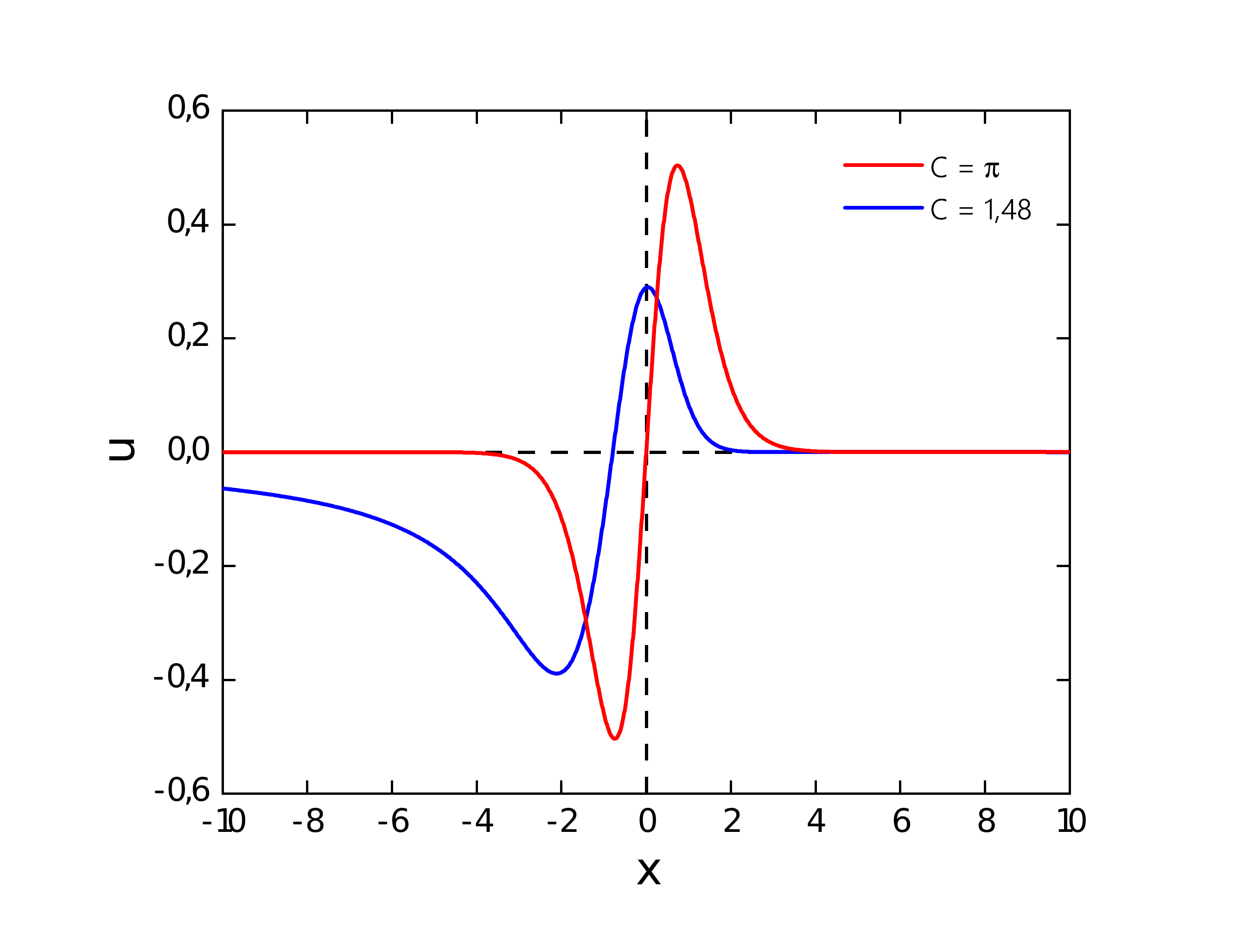}
    \includegraphics[trim=50 30 70 30,clip,width=0.4\textwidth]
        {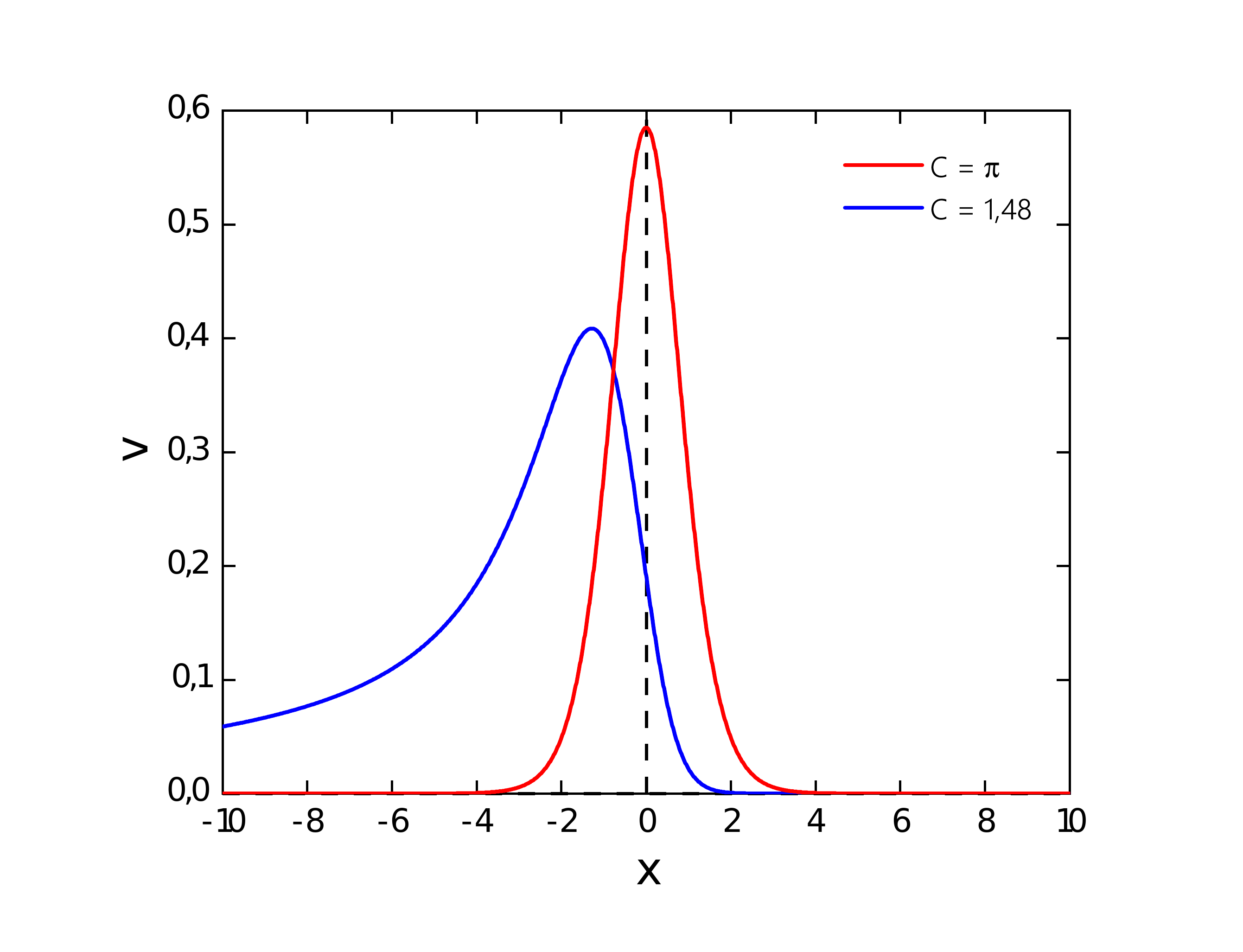}
    \end{center}
    \vspace*{0mm}
    \caption{\small{Components of the localized fermionic modes of the types $A_0$ (upper plot), $A_1$ (middle row) and $B_1$ (bottom row)
    are plotted as functions of the coordinate $x$ for $m=0$, $g=1$ and some set of values of the Yukawa coupling parameter $C$.}}
    \label{fig1a}
\end{figure}

\begin{figure}[t!]
    \begin{center}
    \includegraphics[trim=10 20 70 30,clip,width=0.7\textwidth]
        {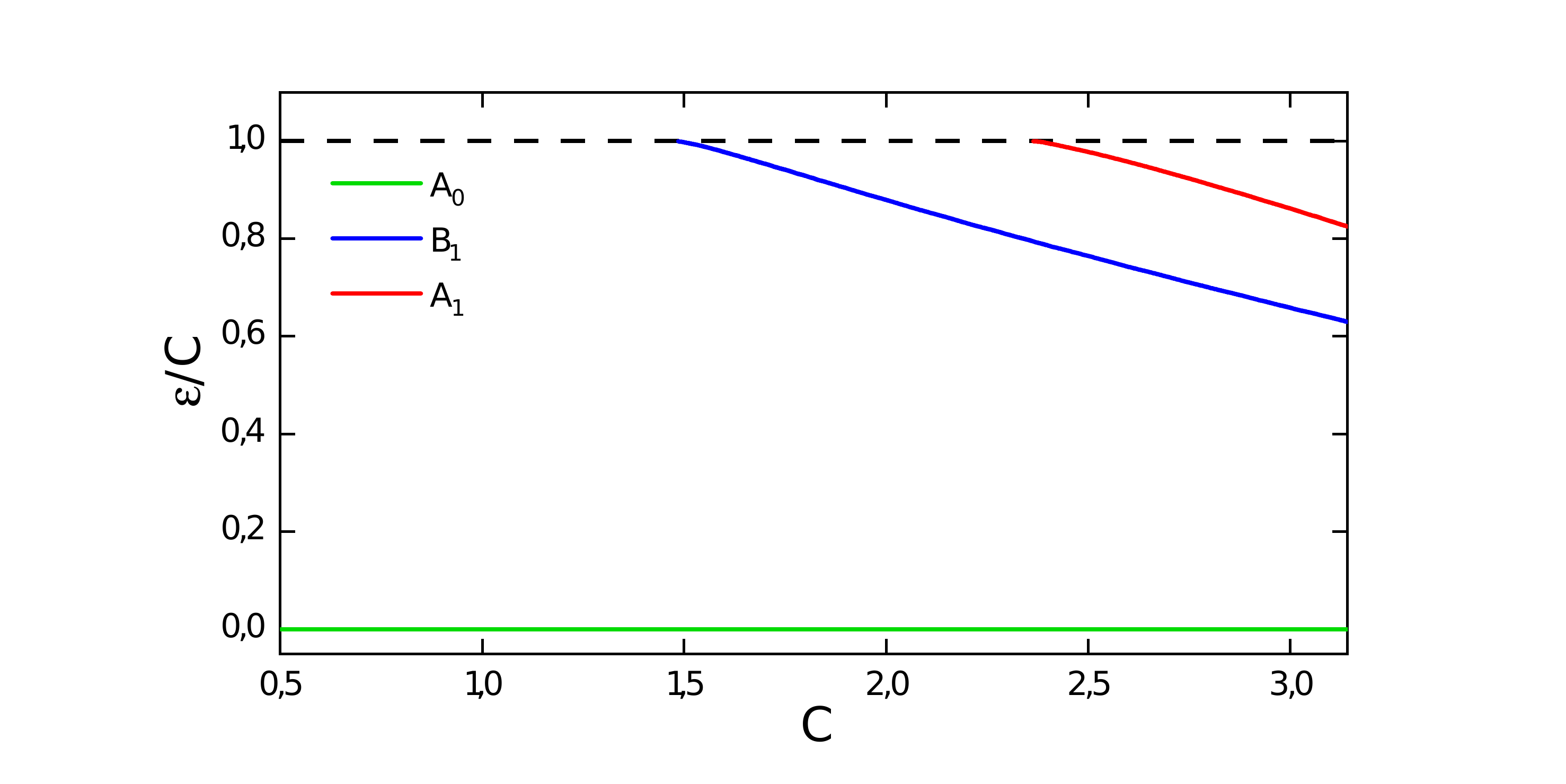}
    \end{center}
    \vspace*{0mm}
    \caption{\small{Normalized energy of the localized fermionic states as a function of the Yukawa coupling parameter $C$
    at $m=0$ and $g=1$ with backreaction. }}
    \label{fig2a}
\end{figure}

Evidently, variation of the Yukawa coupling parameter $C$ affects the fermionic modes localized on the kink. Figure \ref{fig1a} displays the profiles of the corresponding components for some set of values of $C$. Since the effective potential in the equation for the fermions in the external static background field of the kink is no longer symmetric with respect to spatial reflections, the symmetry of the fermionic modes is violated. Our numerical results indicate that the localized asymmetric modes with non-zero eigenvalues may exist for some range of values of $C$. As $C$ deviates from the ``symmetric point'' $C=\pi$, these modes rapidly move towards the continuum, however asymmetric fermionic zero mode $A_0$ exist for all non-zero values of this parameter, see   Fig.~\ref{fig2a}. Indeed, such a mode arise in 1+1 dimensional system according to the index theorem. It disappears at $C=0$.

%%%%%%%%%%%%%%%%%%%%%%%%%%%%%%%%%%%%%%%%%%%%%%%%%%%%%%%%%%%%%%%%%
\section{Kink-fermions collision dynamics}
\label{sec:Collisions}
%%%%%%%%%%%%%%%%%%%%%%%%%%%%%%%%%%%%%%%%%%%%%%%%%%%%%%%%%%%%%%%%%%

Let us now study collision dynamics of a massless fermion state and a sine-Gordon kink in the model \re{lag}.  Hereafter, we consider a  parametrization of the two-component Dirac field which generalizes
the ansatz \re{ansFer} above, which we used for the localized states,
\be \label{TransientAnsatz KinkDirac}
\psi(x,t)=  \left(
\begin{array}{c}
    u(x,t)+i p(x,t)\\
    v(x,t)+i q(x,t)
\end{array}
\right),
\ee
where $u$, $v$, $p$, $q$ are four real functions constrained by the normalization condition $\displaystyle\int\limits_{-\infty}^\infty \left(u^2+p^2+v^2+q^2\right)dx=1$.

Further, in our numerical analysis of the collisions we use a minimal ``non-symmetrized'' Yukawa coupling \re{Yukawa} setting both the bare fermion mass $m=0$ and the coupling parameter $C=0$. This excludes existence of the localized fermion modes, in particular, the zero mode. 

Then, the full system of dynamical equations of the model \re{lag} becomes
\be
\label{TransientEquations KinkDirac}
\begin{split}
    \phi_{xx} + 2g (u v+p q) - \sin\phi &=0\, ,\\
    -u_x-g\phi u&=q_t\, ,\\
    v_x-g\phi v&=p_t\, ,\\
    p_x+g\phi p&=v_t\, ,\\
    -u_x+g\phi q&=u_t\, .\\
\end{split}
\ee
We numerically solve this system of mixed order nonlinear integral-differential equations using the following set of the boundary conditions:
\be
\begin{split}
    u = v = p = q = 0,& \;\; x\to\pm\infty\, ,\\
    \phi = 0,& \;\; x\to -\infty\, , \\
    \phi = 2\pi,& \;\; x\to +\infty\, .
\end{split}
\ee
The initial data used in our simulations represent a normalizable fermionic state of the Gaussian form \cite{Park}:
\be
\label{Fermion_initial}
\psi(x,0)= \frac{1}{\sqrt{2\sqrt{\pi}}}
\exp\left(-\frac{(x-x_0^{})^2}{2}+i k
(x-x_0^{})\right)\left(
\begin{array}{c}
    1\\
    i
\end{array}
\right),
\ee
where $k$ is the wave number and $x_0^{}$ is the initial position of the wave train.

Since we set $m=C=0$, $\mathcal{L}_{\rm int}^{}\to 0$ as $x\to -\infty$, the Gaussian wave packet \re{Fermion_initial} is initially massless and the frequency $\omega =|k|$.  It is coming from the left, i.e.\ it propagates in positive $x$ direction from $x=-\infty$ at the unit speed. As it approaches the kink located at $x=0$, a small dynamical mass of the fermions is generated via the Yukawa coupling with the scalar sector. However, the fermions cannot be localized on the kink.

Note that the presence of the Yukawa interaction term linearly coupled to the scalar sector, mimics the deformations of the scalar potential which induces a force acting on the kink \cite{Kiselev:1998gg}, however this effect is almost negligible. On the other hand, in such a case the field of the sine-Gordon kink generates a potential barrier for the propagating fermions, the probability of transition of the fermions through the barrier is strongly suppressed. Moreover, the mass of the coupled fermions is different in asymptotic states, the modes in  the vacuum $\phi_0^{}=2\pi$ at $x\to +\infty$ have mass $2g\pi$. 

In our calculations, we typically set the initial separation between the sine-Gordon kink, which initially at rest at the center of collision $x=0$, and the incoming train, to $x_0^{}\sim 40-60$ units of length. Interestingly, for relatively small values of the wave number $k$, and $g\le 1$ the dispersion of the normalized propagating relativistic wave train \re{Fermion_initial} is minimal \cite{Park}, the Gaussian packet preserves its form up to collision with the kink for a wide range of values of the parameters of the model. 

\begin{figure}[t!]
    \begin{center}
        \includegraphics[width=0.55\textwidth]
        {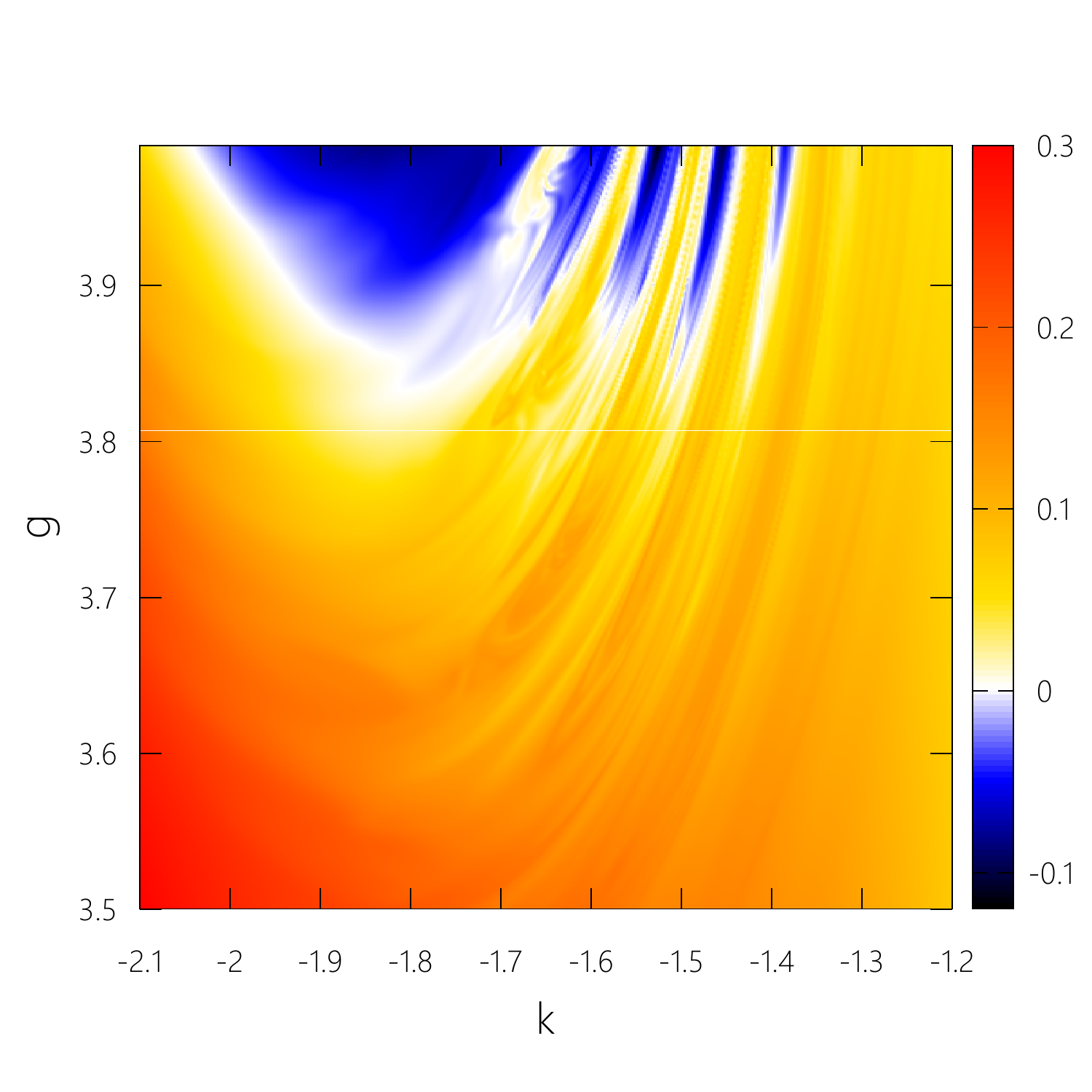}
    \end{center}
    \vspace*{0mm}
    \caption{\small{Post-collision velocity of the kink as a function of the Yukawa coupling constant $g$ and wave number $k$. The initial separation is $x_0 = -45$.}}
    \label{KinkDirac_Speed}
\end{figure}

Figure \ref{KinkDirac_Speed} displays our results  for collisions of the Gaussian train \re{Fermion_initial} with the sine-Gordon kink for some range of values of the parameters $g$ and $k$. Initial separation between the kink and incoming train is fixed as $x_0=-45$.
The figure shows the post-collision velocity of the kink  measured 100 units of time after the impact. 

\begin{figure}[!h]
    \begin{center}
    \includegraphics[width=0.4\textwidth]
        {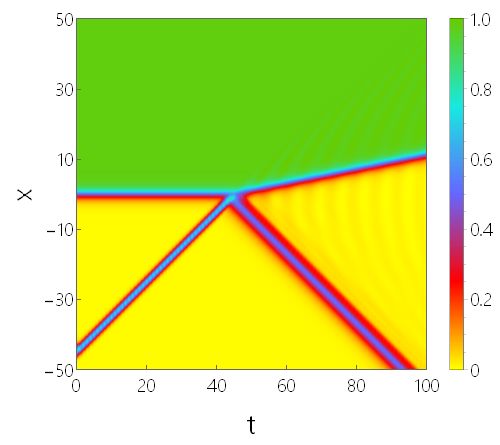}
        \includegraphics[width=0.4\textwidth]
        {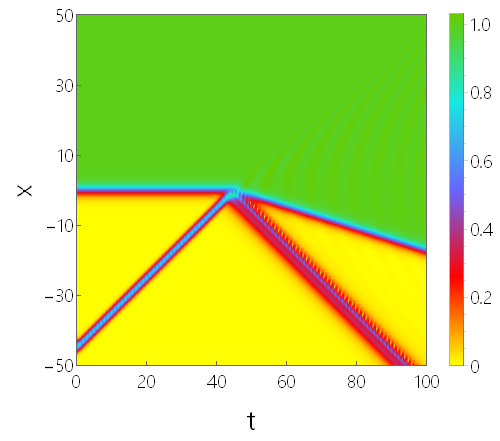}
        \includegraphics[width=0.4\textwidth]
        {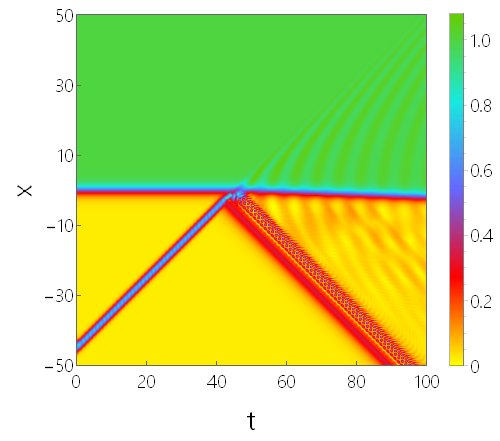}
        \includegraphics[width=0.4\textwidth]
        {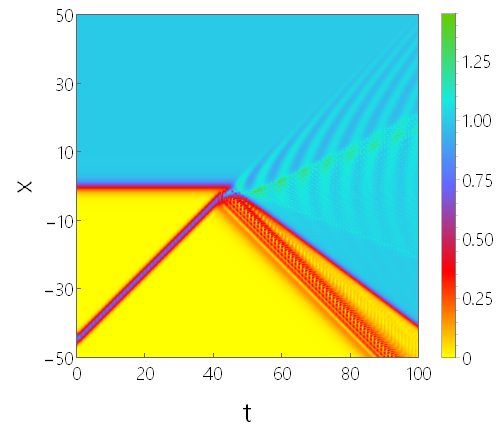}
    \end{center}
    \vspace*{0mm}
    \caption{\small{Collisions between the Gaussian fermionic train \re{Fermion_initial} and the sine-Gordon kink. The plots represent the profiles of the fermion probability density $\bar{\psi}\psi$ and the field of the kink $\frac{\phi}{2\pi}$ as functions of the time $t$ for $g=1$ (upper row) and $g=3$ (bottom row) for the cases of negative (left column) and positive (right column) phases of the train at the moment of collision.}}
    \label{KinkDirac_Profile}
\end{figure}

First, we found that, for sufficiently small values of the coupling constant $g$, the collision dynamics of the system is almost elastic, the fermion train collides with the kink and bounces back providing small transfer of energy to the kink translational mode, as expected. Surprisingly, we observe that the kink is pushed forward by the incident train only if the phase of the fermion packet at the moment of collision is negative, as seen in Fig.~\ref{KinkDirac_Profile}. In the opposite case, the kink starts to move towards the incoming fermion train, resembling the effect of negative radiation pressure in the scalar sector of the $\phi^4$ theory \cite{Romanczukiewicz:2003tn,Forgacs:2008az}. Numerical evaluation shows that, as the coupling constant $g$ remains relatively small, the reflected train almost preserves its Gaussian form with minimal dispersion of the outgoing packet, see Fig.~\ref{KinkDirac_Profile}, upper row. 

The change of the sign of the force, acting on the kink, is related with excitation in the collision of the scalar continuum modes from both sides of the kink. However, if the phase of the fermion train at the impact is positive, a momentum surplus is created behind the kink due to suppression of the scalar radiation outcoming from the left side of the kink in the vacuum sector $\phi_0^{}=0$, and resonance excitation of the continuum modes of the sector $\phi_0^{}=2\pi$ from the right. The situation is opposite, as the phase of the fermion train colliding with the kink is negative\footnote{It was noted that similar effect can be observed in the collision of the incoming Gaussian-like scalar wave trains and the global vortex (W.~Zakrzewski, unpublished note and T.~Romanczukiewicz, private communication.)}.

As $g$ increases beyond 1, the collision dynamics becomes more complicated because of the increasing role of the scalar radiation. First, the amount of energy transferred to the continuum modes in the collision process becomes larger. For some values of the wave number $k$ and the coupling $g$, the kink remains at the center of collision, with scalar radiation outgoing in both directions, see Fig.~\ref{KinkDirac_Profile}, bottom left plot. Moreover, an oscillon can be ejected from the center of collision, see Fig.~\ref{KinkDirac_Profile}, bottom right plot. This chaotic structure can be seen on the upper half of Fig.~\ref{KinkDirac_Speed}. This pattern  is  similar to the behavior of the colliding kinks in the deformed sine-Gordon model \cite{Dorey:2021mdh}.

The production of oscillon is a non-perturbative process with a large energy transfer, i.e., this oscillon carries away a lot of energy and momentum. As a result of the recoil, the kink starts to move quickly. But nevertheless, the kink is still repelled from the fermion packet, fermions cannot form a bound state with the kink. Interestingly, the fermionic packet cannot pass through the kink. This is due to the fact that it is massless to the left of the kink, and it is massive to the right. It is also interesting that the fermionic packet is almost not subject to dispersion. After the collision, there is a small frequency spread, but the fermionic field components do not separate much, they move almost at the same velocity.

Finally, note that further increase of the Yukawa coupling may produce a massive spinor modes propagating on the right side of the excited kink after the impact. This is an effect of the third order, it is related with excitation of the internal mode of the kink and subsequent excitation of the fermionic modes of the continuum. However, the corresponding amplitude, which can be evaluated from the overlap integral between these states, is exponentially suppressed.        

It should also be noted that the problem of scattering of Dirac fermions on the kink was considered in a recent paper \cite{Loginov_2} in a different setup. A modified form of the non-minimal spin-isospin coupling between the kink and the spinor field used in \cite{Loginov_2}  reduces the problem to the modified problem of  scattering of the fermions on the soliton of the non-linear $O(3)$ sigma model. As a results, the field equation for the fermions interacting with the kink can be transformed to a hypergeometric-type equation. The study of the resulting system in paper \cite{Loginov_2}  is simplified further by assumption of smallness of the coupling constant.

%%%%%%%%%%%%%%%%%%%%%%%%%%%%%%%%%%%%%%%%%%%%%%%%%%%%%%%%%%%%%%%%%
\section{Conclusion}
\label{sec:Conclusion}
%%%%%%%%%%%%%%%%%%%%%%%%%%%%%%%%%%%%%%%%%%%%%%%%%%%%%%%%%%%%%%%%%%

In this paper we have studied the fermions interacting with sine-Gordon kink. We discussed the fermion modes, localized on the kink and investigated the effects of backreaction of the fermions on the scalar field.

Our numerical simulations of the collisions between the incoming fermionic train and the sine-Gordon kink revealed a rich diversity of chaotic behaviours mostly related with excitation of the scalar radiation in the process of collision. We have shown that, depending on the phase of the incoming train, the kink can be pulled towards the incident fermion packet, or it can be pushed in the direction of the incoming momentum. This phenomenon resembles the effect of negative radiation pressure in scalar $\phi^4$ model. However, in the case of the fermions interacting with the sine-Gordon kink, the outcome of the collision depends on the wave number of the train. A better understanding of this feature will require a perturbative evaluation of the force acting on the kink after the impact, and subsequent excitation of the continuum scalar modes. However, as things stand, this is a numerical observation.

Increase of the Yukawa coupling strength leads to more complicated pattern of collision dynamics, in particular, we observe production of oscillons and increasing role of the scalar radiation on both sides of the kink. However, massless fermions are always reflected from the kink, it generates a very high potential barrier which forbids the passage.

We  expect that similar effects will be observed in other systems, like $\phi^4$ and $\phi^6$ models coupled to fermions, but we leave the corresponding investigation for future work.

\section*{Note added}

While our paper was posted on the arXiv, the relevant paper \cite{Campos:2022flw} appeared, where the authors consider the kink-antikink collisions in supersymmetric $\phi^4$ theory with fermion back-reaction. This system possesses much larger symmetry that our model, it allows the authors of \cite{Campos:2022flw} to evaluate the fermionic field contribution to the net force, acting on the kink. Interestingly, this force can be both  attractive or repulsive.

\section*{Acknowledgments}

We thank Sergey Fedoruk, Boris Malomed, Irina Pirozhenko and Tomasz Roma\'nczukiewicz for useful discussions. This work was supported by NRNU MEPhI within the Program ``Priority-2030'' under the contract No.~075-15-2021-1305. Ya.S.\ gratefully acknowledges the support by the Ministry of Science and Higher Education of the Russian Federation, project No.~FEWF-2020-003.

\end{document}